\documentclass[fleqn,usenatbib]{mnras}

\usepackage{mathptmx}
\usepackage{txfonts}

\usepackage[T1]{fontenc}
\usepackage{ae,aecompl}

\usepackage{graphicx}	
\usepackage{amssymb}	
\usepackage{pdflscape}
\usepackage{afterpage}
\usepackage{multirow}
\usepackage{longtable}
\usepackage{comment}
\usepackage{caption}
\usepackage{subcaption}
\newcommand{\mjyb}{\,mJy beam$^{-1}$}	
\newcommand{\eps}{\,erg s$^{-1}$ }		

\graphicspath{.{figs/}}

\title[GMRT Galaxy Cluster Key Project]{Imaging results from the legacy GMRT Galaxy Cluster Key Project}

\author[L. T. George et al.]{
Lijo T. George$^{1}$\thanks{E-mail: lijo7george@gmail.com},
Ruta Kale$^{1}$, and,
Yogesh Wadadekar$^{1}$
\\
$^{1}$National Centre for Radio Astrophysics--Tata Institute of Fundamental Research, 
Ganeshkhind, Pune, Maharashtra, INDIA
}

\date{Accepted 2021 August 6. Received 2021 August 6; in original form 2021 May 18}

\pubyear{2019}

\begin{document}
\label{firstpage}
\pagerange{\pageref{firstpage}--\pageref{lastpage}}
\maketitle

\begin{abstract}
	We have used archival GMRT data to image and study 39 galaxy clusters.
	These observations were made as part of the GMRT Key Project on galaxy clusters
	between 2001 and 2004.
	The observations presented in this sample include 14 observations at 610 MHz, 29 at 325 MHz and 3 at 244 MHz
	covering a redshift range of 0.02 to 0.62.
	Multi-frequency observations were made for 8 clusters.
	We analysed the clusters using the SPAM processing software
	and detected the presence of radio halo emission for the first time in the clusters
	RXC J0510-0801 and RXC J2211.7-0349.
	We also confirmed the presence of extended emission in 11 clusters which were known from the literature.
	In clusters where halos were not detected upper limits were placed using our own semi-automated program.
	We plot our detections and non-detections on the empirical $L_X-P_{1.4}$ and $M_{500}-P_{1.4}$ relation 
	in radio halo clusters and discuss the results.
	The best fits follow a power law of the form $L_{500} \propto P_{1.4}^{1.82}$ 
	and $M_{500} \propto P_{1.4}^{3.001}$ which is in accordance with the best estimates in 
	the literature.
\end{abstract}




\section{Introduction}
	Galaxy clusters are the largest gravitationally bound objects in the Universe.
	Diffuse (surface brightness $\sim$mJy arcmin$^{-2}$ at 1.4 GHz), extended ($\sim$Mpc) radio emission 
	from galaxy clusters is seen in the form of \textit{radio halos} and \textit{radio relics}
	\citep[see][for reviews]{feretti12,brujones14,vanweeren19}.
	This is emission from the gas in the Intra Cluster Medium (ICM) and not associated with any galaxies.
	Radio halos are usually located near the centre of the cluster coincident with
	the X-ray distribution of the cluster while radio relics are usually located
	near the edge of the X-ray distribution towards the periphery of the cluster.
	Halos are generally smooth and circular in morphology whereas relics are usually
	elongated and arc-shaped.
	
	\noindent
	Statistically, radio halos and relics are not very common in the Universe.
	Only around a third of the brightest, most massive galaxy clusters ($L_{\rm X}>5\times10^{45}$\eps)
	have been found to host halos and relics \citep{giovannini99,venturi07,venturi08,kale13,kale15}.
	Furthermore, the galaxy clusters also need to be highly disturbed (i.e. show signs 
	of recent merger activity) for increased likelihood to detect the presence of a halo 
	\citep{buote01,cassano10b}.

	\noindent
	The physical origins of both radio halos and relics are still not completely understood.
	While it was earlier suggested that relics trace the outgoing shocks produced during a cluster merger
	and thus accelerate the ICM electrons to relativistic energies via Diffusive Shock Acceleration
	\citep{blandford87,jones91,ensslin98}, recent studies suggest that such a mechanism 
	would not be sufficiently efficient in accelerating electrons to the energies required
	\citep{vazza14,vazza15,vazza16}.
	The models to explain radio halo emission fall under two categories - turbulent acceleration (primary)
	or hadronic (secondary).
	The primary model suggests that the electrons in the ICM are accelerated to relativistic levels
	due to the turbulence generated in a cluster merger \citep{brunetti01,petrosian01}.
	The fact that the majority of radio halos have been discovered in merging galaxy clusters 
	seems to support this model.
	On the other hand, the secondary model predicts that relativistic electrons in galaxy clusters
	are a byproduct of the inelastic collisions between thermal and relativistic protons
	\citep{dennison80,blacola99,pfrosslin04}.
	The biggest shortcoming of this model is that it also predicts gamma rays to be produced via the 
	same process but so far no such emission has been detected 
	\citep{aharonian09a,aharonian09b,ackermann10,aleksic10}.
	It is also suggested that electrons produced via the secondary model could be as much as 10 times
	weaker than those produced via the primary model \citep{brulaz11}.
	It should be noted that the secondary model is not restricted by the dynamical state of the cluster.

	\noindent
	It has also been observed that radio halos exhibit an empirical correlation between their radio power
	at 1.4 GHz ($P_{1.4}$) and the X-ray luminosity ($L_X$) of their host galaxy cluster 
	\citep{feretti00,liang2000,govoni01,bacchi03,brunetti07,brunetti09}.
	Since most halos have been detected in luminous galaxy clusters, this relation tends to hold true 
	in that regime.
	However, if the secondary model is to be believed then there must also exist radio halos 
	with weak radio power that the current generation of radio telescopes are not sensitive enough
	to detect.
	The next generation of radio interferometric telescopes like the Murchison Widefield Array (MWA)
	\citep{lonsdale09,bowman13,tingay13}, Australian SKA Pathfinder (ASKAP) \citep{johnston09}, LOw Frequency
	ARray (LOFAR) \citep{vanhaarlem13} and the upcoming Square Kilometer Array (SKA) \citep{dewdney13}
	are expected to improve in low surface brightness sensitivity and uv-coverage as compared to 
	the existing telescopes.
	It is possible then that many new low radio power radio halos could be detected which will 
	shed new light on the science behind their formation.

	\noindent
	The Giant Metrewave Radio Telescope (GMRT) has been at the forefront of the study of radio halos 
	and relics at low frequencies ($<1$GHz).
	The GMRT Radio Halo Survey (GRHS) \citep{venturi07,venturi08} and its follow up the extended 
	GMRT Radio Halo Survey (eGRHS) \citep{kale13,kale15} detected several radio halos, relics as well as
	mini-halos in a flux limited cluster sample. 
	They also placed upper limits to halo emission in clusters where no halo was detected.
	These upper limits suggest a \textit{bi-modal} nature of galaxy clusters in the $L_X-P_{1.4}$ plot.
	While it is possible that these clusters truly do not contain any halos it is also possible
	that their detection is simply limited by the sensitivity limit of the GMRT. 
	However, even before the (e)GRHS there was another attempt to observe and study extended emission
	from galaxy clusters with the GMRT.

\subsection{GMRT Cluster Key Project}
	The GMRT Key Project was initiated by V. K. Kulkarni and Gopal-Krishna in 2001
	with the aim of imaging a well-defined sample of galaxy clusters to search
	for sources of diffuse radio emission in them.
	Over the course of 5 GMRT cycles, 39 clusters were observed with the GMRT
	at 244, 325 and 610 MHz.
	There were 14 clusters observed at 610 MHz and only 3 observed at 244 MHz.
	The majority of the clusters (29) were observed at 325 MHz.
	Multi-frequency observations (2 or more frequency bands) were available
	for only 8 clusters.
	The aim when creating the sample was to map the parameter space across
	three axes: halo/relic radio luminosity, X-ray luminosity, and a quantifier 
	of the cluster's dynamical state (inferred from the X-ray distribution).
	Furthermore, the project was to be complemented with X-ray observations
	from the \textit{Chandra} and \textit{XMM-Newton} telescopes which could be used 
	to estimate the X-ray luminosity/temperature of the cluster and also quantify the
	dynamical state of the cluster.

	\noindent
	The clusters in the sample ranged in redshift from 0.02 to 0.62.
	The motivation to observe clusters at moderate redshifts ($z>0.3$)
	was to look for evolutionary trends in the radio properties since merger
	activity is expected to be higher at intermediate redshifts.
	
\begin{figure}
		\includegraphics[width=\columnwidth]{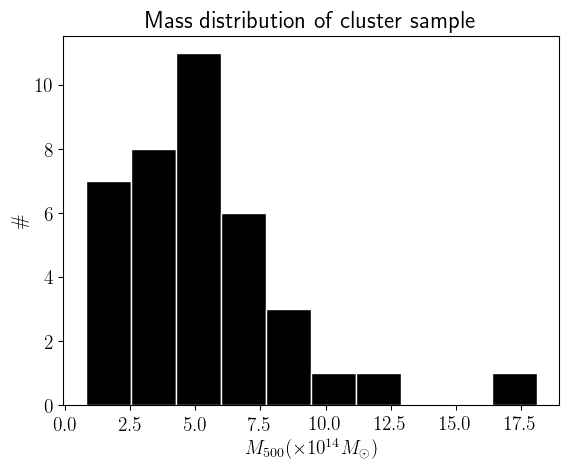}
		\caption{Histogram of the cluster mass ($M_{500}$) of all the clusters in our sample.}
		\label{fig:hist_mass}
	\end{figure}

	\noindent
	Figure~\ref{fig:hist_mass} shows the mass distribution of all the clusters
	in our sample. The masses have been obtained from the Meta-Catalogue of X-ray Clusters
	(MCXC, \cite{piffaretti11}). All the masses were obtained from various surveys
	based on the ROSAT All-Sky Survey (RASS, \cite{truemper93}).
	The luminosities and masses of the clusters in the various surveys
	were homogenised to $L_{500}$ and $M_{500}$	by the authors.
	In our sample, the $L_{500}$ ranges from $0.22\times10^{44}$ \eps to 
	$24\times10^{44}$ \eps. We that no values could be found for the cluster 
	RX J1046.8-2535.
	Note that $L_{500}$ and $M_{500}$ are the X-ray luminosity and mass of the cluster within
	the radius $R_{500}$ of the cluster, where $R_{500}$ is the radius at which the cluster
	mass density is 500 times the critical mass density of the Universe.
	The masses in particular were calculated from the X-ray luminosities using the 
	luminosity mass relation for galaxy clusters as provided by \cite{arnaud10}.
	
\begin{figure}
		\includegraphics[width=\columnwidth]{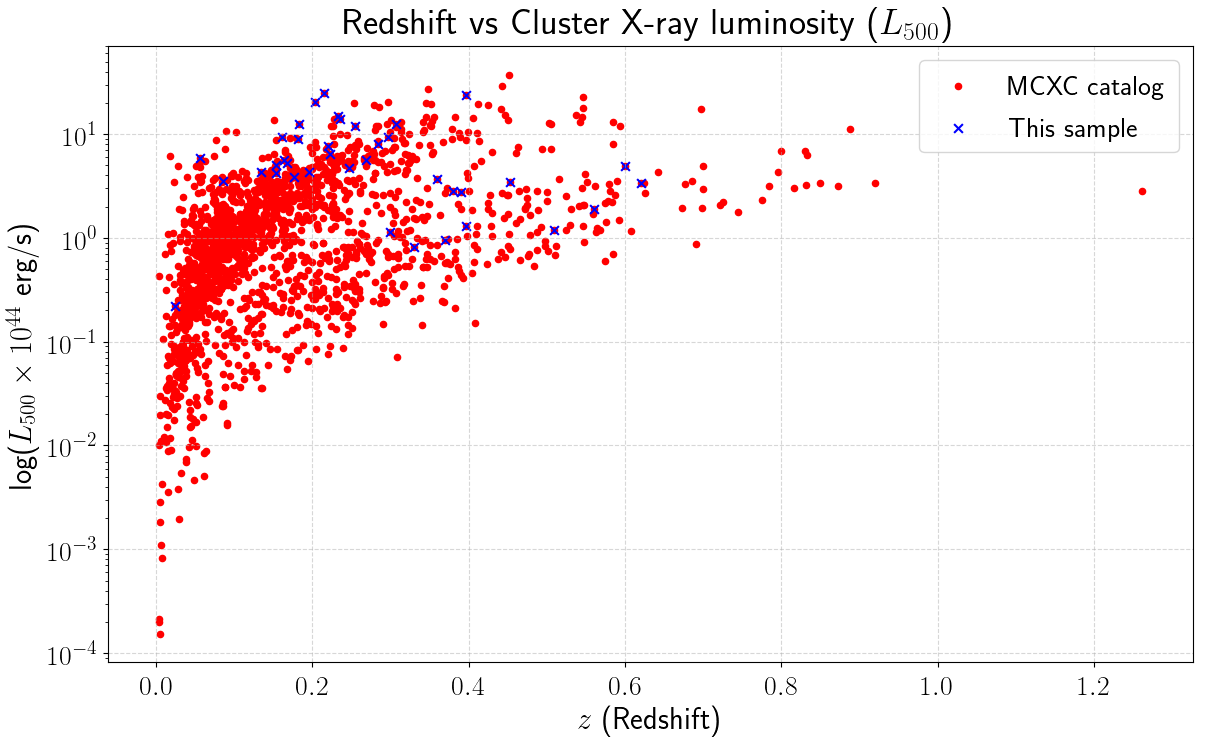}
		\caption{X-ray luminosity ($L_{500}$) vs redshift plot of all the clusters in the MCXC (red)
		along with all the clusters in our sample (blue).}
		\label{fig:l-z}
	\end{figure}

	In Figure~\ref{fig:l-z} we also compare all the clusters
	in our sample with all other clusters in the MCXC in an X-ray luminosity vs redshift plot.

	\noindent
	In Section \ref{sec:obs} we give the details about the observations and 
	the software used for our analysis.
	In Section \ref{sec:results} we briefly discuss the results of our analysis
	including two new detections of radio halos.
	Finally, we end with Sections \ref{sec:discuss} and \ref{sec:conc} with a discussion 
	on our results including how non-detections were handled and the main conclusions of this paper.
	The cosmology used in this paper is as follows:
	$\Omega_0 = 0.3, \Omega_{\Lambda} = 0.7, H_0 = 68$ km s$^{-1}$ Mpc$^{-1}$.

\section{Observations and Analysis}
\label{sec:obs}
	The 39 clusters in this sample were observed between 11 January 2002 and
	25 April 2004 which corresponds to Cycle 1 to Cycle 5 of the GMRT.
	Table \ref{tab:datetime} shows the list of clusters and the date they
	were observed along with the time spent on source at each frequency.
	These observations used the old GMRT hardware correlator as the backend.

	\noindent
	We used the software \textsc{SPAM} \citep{intema14, intema17} to analyse the data initially.
	Subsequent analysis that required continuum uv-subtraction to image the 
	extended emission in the galaxy cluster were performed using the \textsc{CASA} software package
	\citep{mcmullin07}.

	\subsection{SPAM}
	SPAM (Source Peeling and Atmospheric Modeling) is a radio astronomy data processing software
	developed by Huib Intema. 
	The software uses Parseltongue \citep{parseltongue12} and Obit \citep{obit13} as wrappers around 
	the well-known software AIPS (Astronomical Image Processing Software) \citep{wells85} 
	to provide a python interface for radio data processing.
	Several python packages like numpy and scipy are used in scripts that call on 
	Parseltongue and subsequently AIPS for analysis.
	
	\noindent
	The software uses the flux calibrator as the primary source for calibration
	and uses it to perform both flux and phase calibration on the target source.
	RFI (Radio Frequency Interference) mitigation is performed using a combination of
	outlier removal along the time and channel axes as well as modelling and subtration of 
	low level RFI and excision of particulary high amplitude data 
	in the uv-plane with greater weights.
	Both direction-independent as well as direction-dependent calibration are performed 
	by the software. 

	\noindent
	The raw visibility file is provided to SPAM which performs
	an initial calibration (using the primary calibrator) and RFI excision from the 
	visibilities to produce a calibrated visibility file for all sources.
	In the next step, the target visibility file is processed while also undergoing further  
	flagging and calibration (including self-calibration) during the process.
	The imaging parameters are provided at this stage.
	For our analysis, since we are interested in extended emission we used the \textit{briggs} 
	weighting scheme with a \textit{robust} value of 0. 
	This provides a decent compromise between sensitivity and resolution that we require.
	The final image produced at the end of this process is used for analysis.
	All other SPAM configuration parameters were kept at their default values.
	
	\noindent
	Since all of the data used in this work was recorded with the old GMRT correlator 
	an additional step was sometimes required at the beginning of the process. 
	In cases where the data was split into the upper side-band (USB) and lower side-band (LSB)
	the data had to be combined before processing it through SPAM.
	This is also done using the \texttt{combine\_usb\_lsb} task in SPAM.

	\begin{table}
	\centering
	\caption{Cluster list with date of observation and time on source}
	\label{tab:datetime}
	\scalebox{0.7}{
		\begin{tabular}{lccc}
		\hline
		Source&	Frequency&	Date of Observation	&	Time on Source\\
			  &  (MHz)	 &	(dd-mm-yyyy)		&	(min)\\
		\hline\hline
		RX J1334.3+5030					&	610	&	11/01/2002	&	417	\\\hline
		\multirow{3}{*}{RX J0505.3-2849}	&	325	&	22/08/2002	&	134	\\ &	610&	07/09/2002&	254\\&	610&	11/01/2002	& 202\\\hline
		\multirow{2}{*}{RX J1241.5+3250}	&	325	&	06/02/2003	&	416	\\ &	610	&	11/01/2002	&	378	\\\hline
		RXJ0256.5+0006						&	610	&	12/01/2002	&	311	\\\hline
		\multirow{2}{*}{RX J0318.2-0301}	&	325	&	21/08/2002	&	316	\\ &	610&	02/05/2002	& 152\\\hline
		\multirow{4}{*}{RX J1200.8-0327}	&	325	&	01/03/2002	&	225	\\ &	325	&	02/03/2002	& 74	\\&	610	&	02/05/2002&	277\\&	610&	02/06/2002	& 169\\\hline
		Abell 2597							&	610	&	03/05/2002	&	336	\\\hline
		\multirow{3}{*}{RX J0426.1+1655}	&	244	&	19/07/2002	&	249	\\ &	325&	02/03/2002	& 261\\&	610&	06/09/2002	&	256\\	&	610	&	02/06/2002	&	274\\\hline
	    \multirow{3}{*}{RX J2237.0-1516}	&	325	&	02/03/2002	&	261	\\ &	610	&	11/01/2002	& 297	\\&	610&	02/06/2002	&	169	\\\hline
        Abell 2319							&	325	&	02/06/2002	&	253	\\\hline
        \multirow{2}{*}{Abell 2163}			&	325	&	29/12/2003	&	380	\\ &	610&	17/07/2002	& 206\\\hline
        \multirow{2}{*}{Abell 2345}			& 610(E)&	06/09/2002	&	41	\\ &	610(E)&	07/09/2002	& 115\\& 610(W)&	06/09/2002	&	41\\ &	610(W)&	07/09/2002	&	162\\\hline
        RX J1701.3+6414					&	325	&	21/08/2002	&	189	\\\hline
        Abell 2390							&	244	&	18/07/2002	&	200	\\\hline
        Abell 665							&	244	&	19/07/2002	&	229	\\\hline
        \multirow{2}{*}{RXC J1308.5+5342}	&	325	&	14/02/2003	&	392	\\ &	610	&	19/12/2002	& 382\\\hline
        \multirow{2}{*}{RX J0847.1+3449}	&	325	&	05/02/2003	&	318	\\ &	610	&	19/12/2002	& 333\\\hline
        Abell 0400							&	325	&	05/02/2003	&	251	\\\hline
        RX J1120.1+4318					&	325	&	13/02/2003	&	361	\\\hline
        Abell 1084							&	325	&	19/07/2003	&	798	\\\hline
        RXC J2211.7-0349					&	325	&	19/07/2003	&	761	\\\hline
        Abell 0545							&	325	&	19/07/2003	&	521	\\\hline
        Abell 1689							&	325	&	20/07/2003	&	413	\\\hline
        RXC J2014.8-2430					&	325	&	20/07/2003	&	497	\\\hline
        Abell 0901							&	325	&	29/12/2003	&	372	\\\hline
        Abell 0291							&	325	&	29/12/2003	&	468	\\\hline
        RXC J1212.3-1816					&	325	&	31/12/2003	&	612	\\\hline
        RXJ1046.8-2535						&	325	&	19/01/2004	&	427	\\\hline
        Abell 2104							&	325	&	19/01/2004	&	418	\\\hline
        RXC J0510.7-0801					&	325	&	19/01/2004	&	499	\\\hline
        AS0780								&	325	&	20/01/2004	&	504	\\\hline
        Abell 2485							&	325	&	19/01/2004	&	513	\\\hline
        RXC J0437.1+0043					&	325	&	25/04/2004	&	402	\\\hline
        Abell 0907							&	325	&	25/04/2004	&	63	\\\hline
        Abell 3444							&	325	&	25/04/2004	&	126	\\\hline
        Abell 1300							&	325	&	25/04/2004	&	252	\\\hline
        RXC J1504.1-0248					&	325	&	25/04/2004	&	282	\\\hline
        RXC J1514.9-1523					&	325	&	25/04/2004	&	292	\\\hline
        Abell 2537							&	325	&	25/04/2004	&	493	\\\hline
		\end{tabular}
		}
	\end{table}

\section{Results}
\label{sec:results}
	Table \ref{tab:clusterlist} contains the list of all clusters as well as some information
	about the clusters and the images produced with SPAM.


	\begin{landscape}
		\begin{table}
		\centering
		\caption{Cluster list.
		Col. 1: Source Name,
		Col. 2: Redshift,
		Col. 3: Right Ascension (J2000),
		Col. 4: Declination (J2000),
		Col. 5: X-ray Luminosity in the [0.1-2.4] keV range$^1$,
		Col. 6: Total mass$^1$,
		Col. 7: Cluster radius$^1$,
		Col. 8: Frequency of observation (MHz),
		Col. 9: RMS of the image,
		Col. 10: Beam size (in arcsec $\times$ arcsec) and position angle (in degree).\\
		$^1$These values and their definitions can be found in the MCXC \protect\citep{piffaretti11}}
		\label{tab:clusterlist}
		\begin{tabular}{lccccccccc}
			\hline
			Source & $z$ & RA & DEC & $L_{\rm 500}$ & $M_{\rm 500}$ & $R_{\rm 500}$ & Frequency & RMS & Beam size \\
			\multicolumn{2}{}{} & hh:mm:ss & dd:mm:ss & $10^{44}$ \eps & $10^{14} \times M_{\odot}$& Mpc &MHz & \mjyb & (arcsec)$\times$(arcsec), $^{\circ}$ \\
			\hline\hline
			RX J1334.3+5030	&	0.62&	13:34:20.4&	50:31:05.02&	3.4056 & 2.6559 & 0.7786& 610&	0.15&	9.70 $\times$ 8.10, -74.99\\
			\hline
			\multirow{2}{*}{RX J0505.3-2849}	&	\multirow{2}{*}{0.509}	&	\multirow{2}{*}{05:05:19.9}	&	\multirow{2}{*}{-28:49:05.2}	&	\multirow{2}{*}{1.1966}& \multirow{2}{*}{1.5420} & \multirow{2}{*}{0.6788} & 325&0.23& 13.44 $\times$ 8.68, -9.69\\&\multicolumn{6}{}{}&610	&0.39 &9.35 $\times$ 6.68, 12.86\\
			\hline
			\multirow{2}{*}{RX J1241.5+3250}	&	\multirow{2}{*}{0.39}	&	\multirow{2}{*}{12:41:33.2}	&	\multirow{2}{*}{32:50:23}	&	\multirow{2}{*}{2.7505}& \multirow{2}{*}{2.8284} & \multirow{2}{*}{0.8704} &325& 0.16& 11.16 $\times$ 9.52, -16.92\\&\multicolumn{6}{}{}&610&	0.10&	5.50 $\times$ 5.10, -59.97\\
			\hline
			RX J0256.5+0006		&	0.36&	02:56:33	&	00:06:12	& 3.6630 & 3.4520 & 0.941&610&	0.18&	12.20 $\times$ 6.00, 88.84\\
			\hline
			\multirow{2}{*}{RX J0318.2-0301}	&	\multirow{2}{*}{0.37}	&	\multirow{2}{*}{03:18:17.5}	&	\multirow{2}{*}{-03:01:14.02}	&	\multirow{2}{*}{0.9629}& \multirow{2}{*}{1.5161} & \multirow{2}{*}{0.7125} & 325&0.29& 11.40 $\times$ 8.30, 35.22\\&\multicolumn{6}{}{}&610	&0.16 &6.90 $\times$ 4.30, 59.23\\
			\hline
			\multirow{2}{*}{RX J1200.8-0327}	&	\multirow{2}{*}{0.396}	&	\multirow{2}{*}{12:00:49.4}	&	\multirow{2}{*}{-03:27:30}	&	\multirow{2}{*}{1.3039}& \multirow{2}{*}{1.7854} & \multirow{2}{*}{0.745} & 325&1.40& 26.22 $\times$ 11.40, 38\\&\multicolumn{6}{}{}&610	&2.21 &6.63 $\times$ 4.73, 3.8\\
			\hline
			Abell 2597	&	0.0852&	23:25:20	&	-12:07:38	& 1.3039 & 1.7854 & 0.745& 610	&0.40	&9.00 $\times$ 6.26, -40.96\\
			\hline
			\multirow{3}{*}{RX J0426.1+1655}	&	\multirow{3}{*}{0.38}	&	\multirow{3}{*}{04:26:07.4}	&	\multirow{3}{*}{16:55:12}	&	\multirow{3}{*}{2.8406} & \multirow{3}{*}{2.9084} & \multirow{3}{*}{0.8820} & 244&1.18& 22.12 $\times$ 10.79, 82.82\\&\multicolumn{6}{}{}&325&0.53& 15.96 $\times$ 8.02, 57.88\\&\multicolumn{6}{}{}&610	&1.48 &5.61 $\times$ 4.05, 64.32\\
			\hline
			\multirow{2}{*}{RX J2237.0-1516}	&	\multirow{2}{*}{0.299}	&	\multirow{2}{*}{22:37:00.7}	&	\multirow{2}{*}{-15:16:08}	&	\multirow{2}{*}{1.1314}& \multirow{2}{*}{1.7713} & \multirow{2}{*}{0.7709} & 325&0.25& 16.45 $\times$ 9.44, 60.86\\&\multicolumn{6}{}{}&610	&0.08 &6.70 $\times$ 4.30, 38.04\\
			\hline
			Abell 2319	&	0.0557&	19:21:08.8	&	43:57:29.99	& 5.9418 & 5.8345 & 1.2483&610&0.06	&5.50 $\times$ 4.60, 29.31\\
			\hline
			\multirow{2}{*}{Abell 2163}	&	\multirow{2}{*}{0.203}&	\multirow{2}{*}{16:15:34.13}&	\multirow{2}{*}{-06:07:26.29}&	\multirow{2}{*}{20.1585}& \multirow{2}{*}{11.0518} & \multirow{2}{*}{1.4697} &	325&	0.38&	13.76 $\times$ 9.55, 44.53\\&\multicolumn{6}{}{}&	610	&0.12&	8.06 $\times$ 4.20, 55.69\\
			\hline
			\multirow{2}{*}{Abell 2345}	&	\multirow{2}{*}{0.176}&	\multirow{2}{*}{21:27:11}&	\multirow{2}{*}{-12:09:33.01}&	\multirow{2}{*}{3.9026} & \multirow{2}{*}{4.1441} & \multirow{2}{*}{1.0699} &	610 (E)&	0.38&	5.93 $\times$ 5.30, 36\\&	\multicolumn{6}{}{}&	610(W)&	0.43&	7.13 $\times$ 4.62, 30\\
			\hline
			RX J1701.3+6414&	0.453&	17:01:08.83	&	64:14:38.4	& 3.4926 & 3.1055 & 0.8763 &325&0.24	&15.77 $\times$ 9.82, -5.18\\
			\hline
			Abell 2390		&	0.2329&	21:53:35.5	&	17:41:12.01	& 14.815 & 8.9525 & 1.3554 & 244&0.81	&18.50 $\times$ 13.68, 57.26\\
			\hline
			Abell 665		&	0.1818&	08:30:45.19	&	65:52:55.31	& 8.9977 & 6.8668 & 1.2635 &	244&	0.74&	19.37 $\times$ 13.84, -6.7\\
			\hline
			\multirow{2}{*}{RXC J1308.5+5342}	&	\multirow{2}{*}{0.33}	&	\multirow{2}{*}{13:08:31.1}	&	\multirow{2}{*}{53:42:06.98}	& \multirow{2}{*}{0.8107} & \multirow{2}{*}{1.4101} & \multirow{2}{*}{0.7062} & 325&0.12& 11.40 $\times$ 8.30, 35.22\\&\multicolumn{6}{}{}&610	&0.05 &6.3 $\times$ 4.7, 37.65\\
			\hline
			\multirow{2}{*}{RX J0847.1+3449}	&	\multirow{2}{*}{0.56}	&	\multirow{2}{*}{08:47:11.3}	&	\multirow{2}{*}{34:49:16}	&	\multirow{2}{*}{1.1916}& \multirow{2}{*}{1.9683} & \multirow{2}{*}{0.7216} & 325&0.10& 10.06 $\times$ 8.70, 26.6\\&\multicolumn{6}{}{}&610	&0.06 &6.9 $\times$ 4.7, 57.86\\
			\hline
			Abell 0400		&	0.0238&	02:56:30	&	06:10:00	& 0.2211 & 0.8012 & 0.6505&325& 0.84& 10.01 $\times$ 9.50, 43.67\\
			\hline
			RX J1120.1+4318&	0.6	&	11:20:07.38	&	43:18:07.16	& 4.9580 & 3.3968 & 0.8519&325&	0.17&	11.60 $\times$ 8.90, 28.99\\
			\hline
			Abell 1084		&	0.1342&	10:44:33	&	-07:04:22	& 4.2956 & 4.5308 & 1.1182	&325&	0.20&	15.80 $\times$ 8.80, 62.54\\
			\hline
		\end{tabular}
		\end{table}
	\end{landscape}
	\addtocounter{table}{-1}
	\begin{landscape}
		\begin{table}
		\centering
		\caption{(continued) Cluster list }
		\begin{tabular}{lccccccccc}
			\hline
			Source & $z$ & RA & DEC & $L_{\rm 500}$ & $M_{\rm 500}$ & $R_{\rm 500}$ &Frequency & RMS & Beam size \\
			\multicolumn{2}{}{} & hh:mm:ss & dd:mm:ss & $10^{44}$ \eps & $10^{14}\times M_{\odot}$& Mpc & MHz & \mjyb & (arcsec)$\times$(arcsec), $^{\circ}$ \\
			\hline\hline
			RXC J2211.7-0349&	0.397&	22:11:44.6&	-03:49:47&	24.0000 & 18.1000 & 1.6100 &	325&	0.20&	12.80 $\times$ 9.80, 72.06\\
			\hline
			Abell 0545		&	0.154&	05:32:23.1&	-11:31:50.02& 5.0062 & 4.9028 & 1.1403	&	325&	0.32&	12.70 $\times$ 8.80, 30.58 \\
			\hline
			Abell 1689		&	0.1832&	13:11:29.5&	-01:20:17.02 &	12.5240 & 8.3920 & 1.3502	&	325&	0.30&	12.20 $\times$ 8.40, 48.93\\
			\hline
			RXC J2014.8-2430&	0.1612&	20:14:50&	-24:30:35&	9.4586 & 7.1884 & 1.2922	&	325&	-&	-\\
			\hline
			Abell 0901		&	0.1634&	09:56:26.4&	-10:04:12&	5.6296 & 5.2302 & 1.1613	&	325&	0.2&	13.60 $\times$ 8.60, 39.77\\
			\hline
			Abell 0291		&	0.196&	02:01:44.2&	-02:12:02.99& 4.2718 & 4.3137 & 1.0767	&	325&	0.21&	11.30 $\times$ 10.40, -21.51\\
			\hline
			RXC J1212.3-1816&	0.269&	12:12:18.9&	-18:16:43&	5.6441 & 4.8327 & 1.0892	&	325&	0.16&	15.10 $\times$ 9.60, 12.58\\
			\hline
			RXJ1046.8-2535	&	0.243&	10:46:48&	-25:34:59.99&	-	&	-	& - &	325&	0.16&	17.10 $\times$ 8.70, 34.55\\
			\hline
			Abell 2104		&	0.1533&	15:40:07.5&	-03:18:29.02& 4.2260 & 4.4239 & 1.1021	&	325&	0.45&	10.20 $\times$ 8.30, -80.66\\
			\hline
			RXC J0510.7-0801&	0.2195&	05:10:47.91&	-08:01:44.29& 7.7286 & 6.0827 & 1.1974	&	325&	0.26&	14.60 $\times$ 8.50, 53.62\\
			\hline
			AS0780			&	0.2357&	14:59:29.3&	-18:11:12.98& 13.9675 & 8.6180 & 1.3370	&	325&	0.17&	14.30 $\times$ 9.60, -16.44\\
			\hline
			Abell 2485		&	0.2472&	22:48:32.9&	-16:06:23&	4.6638 & 4.3759	& 1.0622 & 325&	0.37&	12.86 $\times$ 9.43, 10.65\\
			\hline
			RXC J0437.1+0043&	0.2842&	04:37:10.1&	00:43:37.99& 8.0950 & 5.9496 & 1.1608	&	325&	0.21&	11.00 $\times$ 9.37, 50.02\\
			\hline
			Abell 0907		&	0.1669&	09:58:22.2&	-11:03:34.99& 5.2999 & 5.0282 & 1.1448	&	325&	0.59&	13.58 $\times$ 9.30, 55.59\\
			\hline
			Abell 3444		&	0.2542&	10:23:50.8&	-27:15:31&	11.9196 & 7.7124 & 1.2798	&	325&	0.46&	13.60 $\times$ 10.90, -10.11\\
			\hline
			Abell 1300		&	0.3075&	11:31:54.4&	-19:55:41.99& 12.4620 & 7.5980 & 1.2485	&	325&	0.72&	16.30 $\times$ 13.10, 30.19\\
			\hline
			RXC J1504.1-0248&	0.2153&	15:04:07.5&	-02:48:15.98& 24.9688 & 12.4750 & 1.5235	&	325&	0.41&	14.30 $\times$ 9.30, 69.06\\
			\hline
			RXC J1514.9-1523&	0.2226&	15:14:58&	-15:23:10.0& 6.4277 & 5.4232 & 1.1511	&	325&	0.26&	12.80 $\times$ 10.70, 59.53\\
			\hline
			Abell 2537		&	0.2966&	23:08:23.2&	-02:11:30.98& 9.3659 & 6.4393 & 1.1864	&	325&	0.25&	13.18 $\times$ 10.14, 82.04\\
			\hline
		\end{tabular}
		\end{table}
		\end{landscape}
	
		\begin{figure*}
			\centering
			\includegraphics[width=0.9\columnwidth]{MCXCJ0510_7-0801_317}
			\includegraphics[width=0.7\columnwidth]{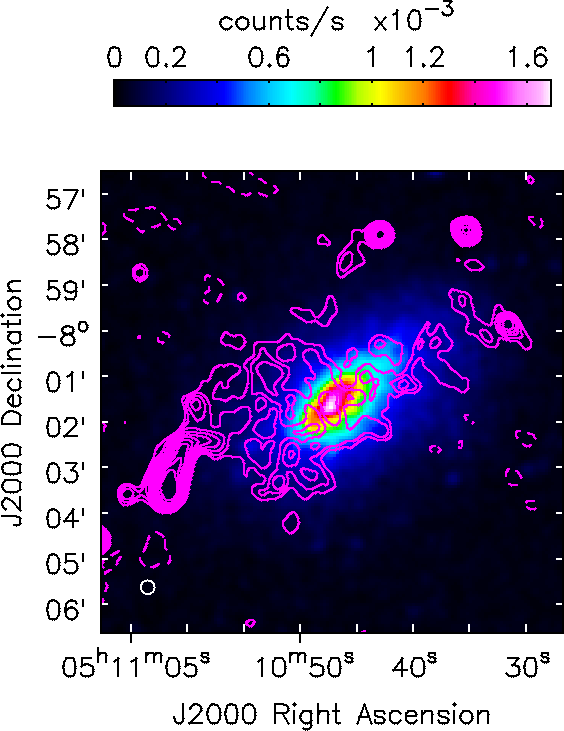}\\
			\includegraphics[width=0.9\columnwidth]{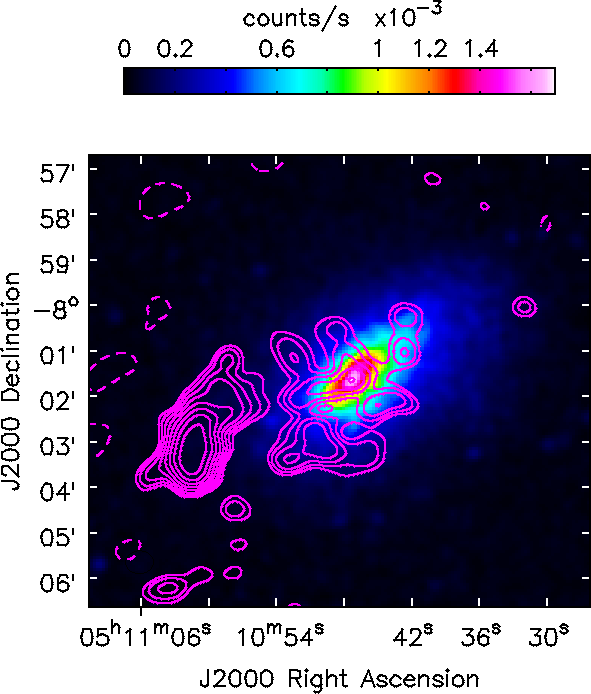}
			\caption{
			{\it Top Left:} Original output image from SPAM for the cluster RXC J0510.7-0801 at 325 MHz in grayscale. 
			The resolution of the image is $14.64''\times8.54'', 53.63^{\circ}$.
			The dotted circle represents the $R_{500}$ for the cluster as shown in the MCXC. 
			The colour bar shows the flux density values in the image in Jy.
			{\it Top Right:} Radio contours for the cluster (magenta) overlaid on the {\it XMM-Newton} 
			X-Ray image. 
			The contours of the radio image start at $3\sigma$, where $\sigma = 0.35$ mJy/beam and increase by $\sqrt{2}$ afterwards.
			The resolution of the image is $18''\times18''$.
			The colour bar above shows the counts per second for the X-ray image.
			{\it Bottom:} Contours of the point source removed and smoothed image of the radio halo in the cluster
			overlaid on the X-ray image.
			The contours of the radio image start at $3\sigma$ and increase by $\sqrt{2}$ afterwards.
			The resolution of the image is $48''\times44'', 45^{\circ}$.
			}
			\label{fig:j0510}
		\end{figure*}

		\begin{figure*}
			\centering
			\includegraphics[width=0.9\columnwidth]{MCXCJ2211_7-0349_317}
			\includegraphics[width=0.7\columnwidth]{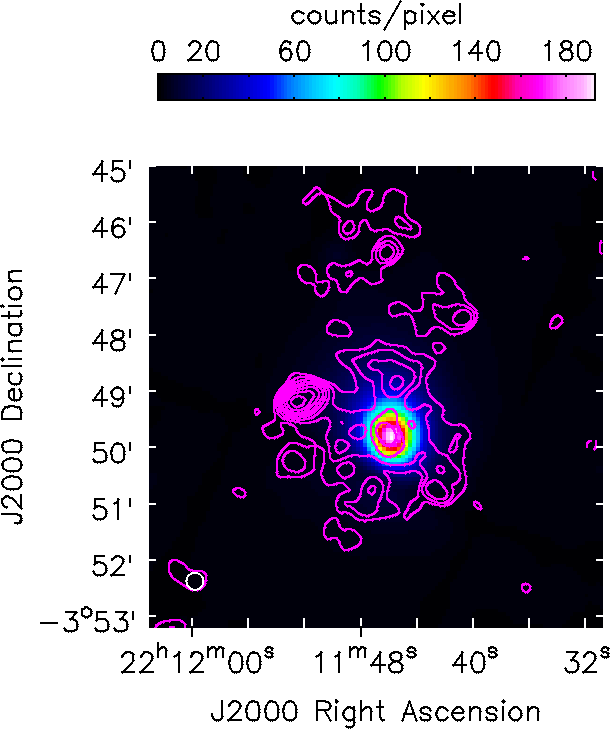}\\
			\includegraphics[width=0.9\columnwidth]{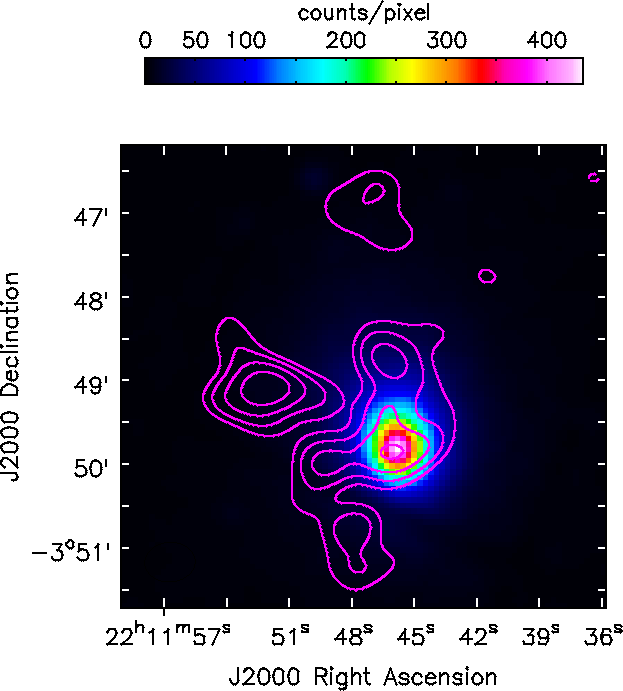}
			\caption{			
			{\it Top Left:} Original output image from SPAM for the cluster RXC J2211.7-0349 at 325 MHz in grayscale. 
			The resolution of the image is $12.8''\times9.48'', 72^{\circ}$.
			The dotted circle represents the $R_{500}$ for the cluster as shown in the MCXC. 
			The colour bar shows the flux density values in the image in Jy.
			{\it Top Right:} Radio contours for the cluster (magenta) overlaid on the {\it XMM-Newton} 
			X-Ray image. 
			The contours of the radio image start at $3\sigma$ where $\sigma=0.4$ mJy/beam and increase by $\sqrt{2}$ afterwards.
			The radio image has been convolved to $18''\times18''$.
			The colour bar above shows the counts per pixel for the X-ray image.
			{\it Bottom:} Contours of the point source removed and smoothed image of the radio halo in the cluster
			overlaid on the X-ray image.
			The contours of the radio image start at $3\sigma$ and increase by $\sqrt{2}$ afterwards.
			The resolution of the image is $36''\times28'', -86^{\circ}$.
			}
			\label{fig:j2211}
		\end{figure*}

	\subsection{New Detections}
		In our sample of 38 clusters we have detected 2 new possible candidate radio halos - 
		RXC J0510.7-0801 and RXC J2211.7-0349.

		\subsubsection{RXC J0510.7-0801}
			This is a massive ($M=7.4\times10^{14}M_{\sun}$) \citep{giacintucci17}, 
			luminous ($L_{\rm X}=12.83\times10^{44}$ \eps \cite{bohringer04}) cluster 
			at a redshift of $z=0.22$ \citep{degrandi99}.
			\cite{kale15} analysed this cluster but could not detect any extended emission as
			the quality of the data were poor.

			Figure \ref{fig:j0510} shows the greyscale radio image of the cluster 
			in the left panel. Also shown is the $R_{500}$ range in the dashed circle.
			The right panel of the figure shows the contours of the radio image 
			overlaid on the X-ray image from \textit{XMM-Newton}.
			The bottom panel of the image shows the point source subtracted image
			smoothed and overlaid on the X-ray image.
			Point source subtraction was performed in CASA by first only imaging the point sources
			by setting the uv-cutoff above $3k\lambda$ and then using the task \textsc{uvsub}
			to remove the point sources from the visibility file. 
			Finally, in order to bring out the extended emission better, imaging was performed
			using only the first $5k\lambda$ using the modified visibility file.

		\subsubsection{RXC J2211.7-0349}
			This is also an extremely massive ($M_{\rm X} = 10.5\times10^{14} M_{\sun}$) 
			and luminous ($L_{\rm X}=15.84\times10^{44}$ \eps \citep{ebeling10}) 
			cluster at a redshift of $z=0.3977$. The gas temperature of the cluster was estimated to be
			$T_{\rm X}=11.3^{+1.46}_{-1.17}$ keV \citep{cavagnolo08}.

			The radio image of this cluster is shown in the left panel of Fig. 
			\ref{fig:j2211}.
			The right panel shows the radio contours on the colour X-ray image.
			The X-ray image is also from the \textit{XMM-Newton} as before.
			The bottom panel of the image shows the point source subtracted image
			smoothed and overlaid on the X-ray image.
			This image was also created using the same method as mentioned above for RXC J0510.7-0801.

			\cite{rossetti16} studied this cluster as part of a \textit{Planck} selected
			sample of clusters to characterize the dynamical state of clusters.
			For this cluster they used results from \cite{whl12} who cross-matched clusters 
			between optical (SDSS) and X-ray (\textit{ROSAT}) catalogs.
			Based on their analysis they estimated the X-ray peak of the cluster to be 
			offset from the BCG in the cluster by $7.16''$ (38.3 kpc).

			In radio the cluster was observed with the GMRT again in 2009 
			(Project ID: 16\_117) at 610 MHz.
			We analysed these data as well using the same method described before.
			However, no sources of extended emission were detected in this image
			(Beam = $6.15''\times4.87'', 43.23^{\circ}$, RMS = 50 \mjyb).

	\subsection{Known sources of diffuse emission}
	Of the 39 clusters in our sample 11 clusters are known in the literature to host radio halos and 
	relics.
	The archival data we used was able to detect all the sources of extended emission in the clusters.
	Table \ref{tab:existing_halos} contains the list of these clusters as well as 
	the flux density values of the sources as estimated by this work compared to the literature.
	It should be noted that difference in flux density values for sources at the same frequency
	could be due to several reasons such as calibration or different sizes of the source
	taken during estimation.
		
		\begin{table*}
			\centering
			\caption{List of clusters with known diffuse emission. 
			The subscript $0$ represents images from this paper
			while the subscript $1$ represents images from the reference paper 
			given in the table.
			Some clusters contain a halo (H) as well as a relic (R) while others contain mini-halos (MH).
			A2345 is the only cluster with relics (East and West) and no central halo.
			}
			\label{tab:existing_halos}
			\scalebox{0.85}{
			\begin{tabular}{lccccr}
				\hline
				Source & $\nu_0$ (MHz) & $S_0$ (mJy) & $\nu_1$ (MHz) & $S_1$ (mJy) & Reference \\
				\hline
				RXC J0256.5+0006 	& 610 & $30.99$ & 610 & $6.9\pm0.7$ & \cite{knowles16,knowles19}\\
				A2163 				& 325 & $1126.09$ (H)/$98$ (R) & 325 & $861\pm10$	& \cite{feretti01,feretti04}		\\
				A2345 				& 610 & $565.17(W)/408.48(E)$ & 325 & $291\pm4 (W)/188\pm3 (E)$ & \cite{bonafede09}		\\
				A665 				& 240 & $462.58$ & 325 & $197\pm6$	& \cite{giovannini00,feretti04}		\\
				A0545			 	& 325 & $624.9$  & 1400 & $23\pm1$ 	& \cite{bacchi03}		\\
				AS0780 				& 325 & $123.3$ (MH)& 610 & $34\pm2$ & \cite{giacintucci19}	\\
				A907 				& 325 & $44.7$ (MH) & 610 & $34.9\pm5.6$ & \cite{giacintucci19}	\\
				A3444			 	& 325 & $21.36$ (MH) & 610 & $10.0\pm0.8$ & \cite{venturi07,giovannini09}		\\
				A1300 				& 325 & $207.67$ (H)/$90.59$ (R) & 325 & $130\pm10$(H)/$75\pm6$(R)	& \cite{reid99, venturi13, parekh17}\\
				RXC J1504.1-0248 	& 325 & $220.8$	(MH) & 325 & $215\pm11$	& \cite{giacintucci11b}\\
				RXC J1514.9-1523 	& 325 & $139.5$ & 325 & $102\pm9$	& \cite{giacintucci11a}\\
				\hline
			\end{tabular}
			}
		\end{table*}
	
	\subsubsection{RX J0256.5+0006}
		The cluster is at a redshift of $z=0.36$ \citep{romer2000}. 
		\cite{plionis05} also give the X-ray luminosity	of the cluster to be 
		$L_{\rm X}=2.84\times10^{44}$ \eps and a temperature of $T_{\rm X} = 5.6^{+0.7}_{-0.5}$.
		Detailed $XMM-Newton$ observations of the cluster by \cite{majerowicz04} revealed 
		that this cluster shows two peaks in X-ray -- 
		one corresponding to the main cluster centre and another to its west --
		which suggests the cluster is in a merger state.
		A study of its dynamics showed that the merger is on-axis 
		and the subcluster is roughly $20-30\%$ of the main cluster by mass.
		
		Recent GMRT 325 MHz and 610 MHz observations of the cluster \citep{knowles16,knowles19} 
		revealed that the cluster does indeed host a faint radio halo.
	
	\subsubsection{Abell 2163}
		A2163 is another massive cluster at a redshift of $z=0.203$ \citep{strood99}.
		It has an X-ray luminosity of $L_{\rm X}=34.37\times10^{44}$ \eps and a temperature of
		$T_{\rm X}=13.4^{+0.45}_{-0.45}$ keV \citep{planck11xi}.
		The cluster hosts a $\sim3$ Mpc size radio halo \citep{herbirk94, feretti01, feretti04}.
	
	\subsubsection{Abell 2345}
		A2345 is an intermediate redshift ($z=0.176$ \cite{strood99}) cluster with an X-ray
		luminosity of $L_{\rm X}=6.47\times10^{44}$ \eps.
		The cluster contains double relics on opposite ends of the cluster \citep{bonafede09}.

	\subsubsection{Abell 665}
		The cluster is at a redshift of $z=0.182$ \citep{strood99} and has an X-ray luminosity of
		$L_{\rm X}=15.17\times10^{44}$ \eps \citep{bohringer00}. The temperature of this cluster
		has been estimated to be $T_{\rm X}=7.5^{+0.2}_{-0.2}$ keV \citep{maughan08}.
		This cluster is well known to host a 1.8 Mpc radio halo \citep{mofbirk89,giovannini00}.

	\subsubsection{Abell 0545}
		A545 has a redshift of $z=0.154$ \citep{strood99} and an X-ray luminosity of
		$L_{\rm X}=8.37\times10^{44}$ \eps \citep{bohringer04} with a temperature of
		$T_{\rm X}=5.5^{+5.5}_{-2.1}$ \citep{david93}.
		The X-ray structure of the cluster is disturbed \citep{buote95, buote01} 
		and radio observations of the cluster \citep{bacchi03} reveal the presence of a giant radio halo.
	
	\subsubsection{AS0780}
		AS0780 or RXCJ1459.4-1811 is a highly luminous cluster $L_{\rm X}=22.94\times10^{44}$ \eps
		found at a redshift of $z=0.236$ \citep{bohringer04}.
		\cite{giacintucci17} first claimed the cluster to host a mini-halo based on VLA 1.4 GHz observations.
		Using new GMRT and VLA observations \cite{giacintucci19} confirmed the presence of the mini-halo.
	
	\subsubsection{Abell 0907}
		A907 has a redshift of $z=0.153$ \citep{ebeling96} and an X-ray luminosity of
		$L_{\rm X}=7.59\times10^{44}$ \eps \citep{bohringer04}. The gas temperature in the cluster has been
		estimated to be $T_{\rm X}=5.3^{+0.1}_{-0.1}$ keV \citep{maughan08}.
		\cite{giacintucci19} claim that the cluster hosts a 65 kpc mini-halo.
	
	\subsubsection{Abell 3444}
		A3444 is an intermediate redshift ($z=0.253$ \cite{strood99}) highly luminous galaxy cluster
		($L_{\rm X}=18.11\times10^{44}$ \eps \cite{bohringer04}) with a temperature of
		$T_{\rm X}=5.6^{+0.24}_{-0.18}$ keV \citep{matsumoto01}.
		X-ray analysis by \cite{lemonon97} suggested that this is a cool-core cluster.
		The GRHS detected a possible extended emission which was confirmed to be a mini-halo 
		recently by \cite{giacintucci19}.
	
	\subsubsection{Abell 1300}
		A1300 is another highly luminous galaxy cluster ($L_{\rm X}=19.23\times10^{44}$ \eps \cite{ebeling10})
		at a redshift of $z=0.307$ \citep{strood99}. The cluster has a temperature of
		$T_{\rm X}=7.75^{+0.31}_{-0.31}$ keV \citep{planck11xi}.
		X-ray and optical observations \citep{lemonon97} revealed that revealed that the cluster
		has a highly disturbed morphology and substructures.
		A radio halo and relic were first detected in this cluster by \cite{reid99}.
		A low frequency follow-up observation with the GMRT was carried out by \cite{giacintucci11} 
		and \cite{venturi13} where the authors detected a faint bridge emission 
		between the halo and relic.
		\cite{parekh17} also performed a multi-frequency analysis of the cluster 
		and confirmed the presence of this bridge as well as a second relic close to the halo.
	
	\subsubsection{RXC J1504.1-0248}
		The cluster has a redshift of $z=0.215$. \textit{Chandra} analysis of the cluster reveals
		the X-ray luminosity of the cluster to be $L_{\rm X} = 2.3\times10^{45}$ \eps \citep{bohringer05}.
		This study revealed the cluster to be the most luminous cluster in the southern sky at redshifts $<0.3$
		with an extremely compact and dense core.
		High frequency radio observations by \cite{mittal09} show that the AGN in the cluster has a flat spectral index ($\sim 0.29$)
		between 1.4 GHz and 4.86 GHz.
		Low frequency observations of the cluster at 327 MHz reveal the presence of a radio mini-halo
		nearly 140 kpc in size \citep{giacintucci11a}.
	
	\subsubsection{RXC J1514.9-1523}
		This cluster has a redshift of $z=0.223$ and an X-ray luminosity of $L_{\rm X}=10.63\times10^{44}$ \eps
		\citep{bohringer04}.
		Radio observations of the cluster show that it contains an ultra steep spectrum radio halo 
		\citep{giacintucci11b}.

	\begin{table}
		\caption{Estimated upper limits to radio halo emission in our sample. 
		Col. 1: Cluster name,
		Col. 2: RMS in the central region,
		Col. 3: Frequency of image,
		Col. 4: Upper limit at this frequency,
		Col. 5: Log of the extrapolated radio power at 1.4 GHz}
		\label{tab:uls}
		\hspace{-0.5cm}
		\scalebox{0.9}{
		\begin{tabular}{lcccc}
			\hline
			Source 	& RMS     & $\nu$ & $S_{\nu}^{\rm UL}$ & log($P_{1.4}^{\rm UL}$)\\
			        & [\mjyb] & [MHz] & [mJy]              & [W Hz$^{-1}$] \\
			\hline
			RX J1334.3+5030 	& 0.14	& 618	&	10.5  	&	24.86 \\
			RX J1241.5+3250 	& 0.165	& 333	&	24.7 	& 	24.38 \\
			RX J0318.2-0301 	& 0.154	& 318	&	13.69 	&	23.65 \\
			RX J1200.8-0327 	& 0.2	& 618	&	1.82	&	22.58 \\
			Abell 2345 			& 0.577 & 618 	&	125.1 	&	24.62 \\
			RX J1701.3+6414 	& 0.284	& 333	&	8.5 	&	24.07 \\
			Abell 2390 			& 0.9	& 240	&	45.2	&	23.92 \\
			RXC J1308.5+5342 	& 0.045	& 618	&	6.00	&	23.33 \\
			RX J0847.1+3449 	& 0.069	& 618	&	11.09	&	23.97 \\
			RX J1120.1+4318 	& 0.181	& 333	&	36.2	&	25.01 \\
			Abell 1084 			& 0.203	& 318	&	101.6	&	23.89 \\
			Abell 1689 			& 0.331	& 317	&	66.12	&	24.01 \\
			Abell 0291 			& 0.264	& 317	&	57.23 	&	24.01 \\
			RXC J1212.3-1816 	& 0.148	& 317	&	13.57	&	23.7  \\
			Abell 2104 			& 0.449	& 317	&	89.8 	&	23.97 \\
			Abell 2485 			& 0.338	& 317	&	16.88	&	23.71 \\
			RXC J0437.1+0043 	& 0.278	& 317	&	55.5 	&	24.37 \\
			Abell 0907 			& 0.634	& 317	&	422 	&	24.72 \\
			Abell 2537 			& 0.254	& 317	&	50.8 	&	24.38 \\
			\hline
		\end{tabular}
		}
	\end{table}

	\begin{table}
		\caption{Clusters with no upper limits calculated.}
		\label{tab:no_uls}
		\hspace{-0.5cm}
		\scalebox{0.95}{
		\begin{tabular}{lcr}
			\hline
			Source 	& $\nu$ (MHz) &	Remarks \\
			\hline
			RX J0505.3-2849	 & 325/610		& Bright source at cluster centre \\
			A2597 			 & 610 			& Bright source at cluster centre \\
			RX J0426.1+1655  & 244/325/610 	& Bright source at cluster centre \\
			RX J2237.0+1516  & 325/610 		& Bad image \\
			A2319 			 & 325 			& Bad image \\
			A0400 			 & 325 			& Bright source at cluster centre \\
			A0901 			 & 325 			& Bad image \\
			\hline
		\end{tabular}
		}
	\end{table}

\section{Discussion}
\label{sec:discuss}
Of the 39 clusters in the original sample we were able to image 38 clusters. 
The data for the cluster RXC J2014.8-2430 was too corrupted for us to properly image. 
The remaining 38 clusters were imaged with the SPAM software package.
In the case of another cluster (RXJ1046.8-2535) no detailed X-ray information was available.

\begin{table}
	\centering
	\caption{Best fit values for $L_X-P_{1.4}$ plot using four methods. 
	The values correspond to a $Y=AX+B$ model where $X = log(L_{500})$ and $Y=log(P_{1.4})$).
	The values in bold are used in the plot.}
	\label{tab:lpfits}
	\begin{tabular}{lcccc}
		\hline
		Method & A & $\sigma_A$ & B & $\sigma_B$\\
		\hline
		\textbf{OLS (Y|X)} & \textbf{1.8190} & \textbf{0.3126} & \textbf{-56.9956} & \textbf{14.0065}\\
		OLS (X|Y) & 4.3765 & 0.9952 & -171.7180& 44.7165\\
		bisector  & 2.6276 & 0.2825 & -93.2691 & 12.7114\\
		orthogonal& 4.0723 & 0.9102 & -158.0726&  40.9044 \\
		\hline
	\end{tabular}
\end{table}

\begin{table}
	\centering
	\caption{Best fit values for $M_X-P_{1.4}$ plt using four methods. 
	The values correspond to a $Y=AX+B$ model where $X = log(M_{500})$ and $Y=log(P_{1.4})$.
	The values in bold are used in the plot.}
	\label{tab:mpfits}
	\begin{tabular}{lcccc}
		\hline
		Method & A & $\sigma_A$ & B & $\sigma_B$\\
		\hline
		\textbf{OLS (Y|X)} & \textbf{3.0095} & \textbf{0.5969} & \textbf{-19.8687} & \textbf{8.8032}\\
		OLS (X|Y) & 8.0905 & 2.1277 & -94.9471 & 31.4991\\
		bisector  & 4.4325 & 0.5095 & -40.8957 & 7.5512\\
		orthogonal& 7.8851 & 2.0587 & -91.9113 & 30.4801\\
		\hline
	\end{tabular}
\end{table}

\begin{figure*}
	\includegraphics[width=\columnwidth]{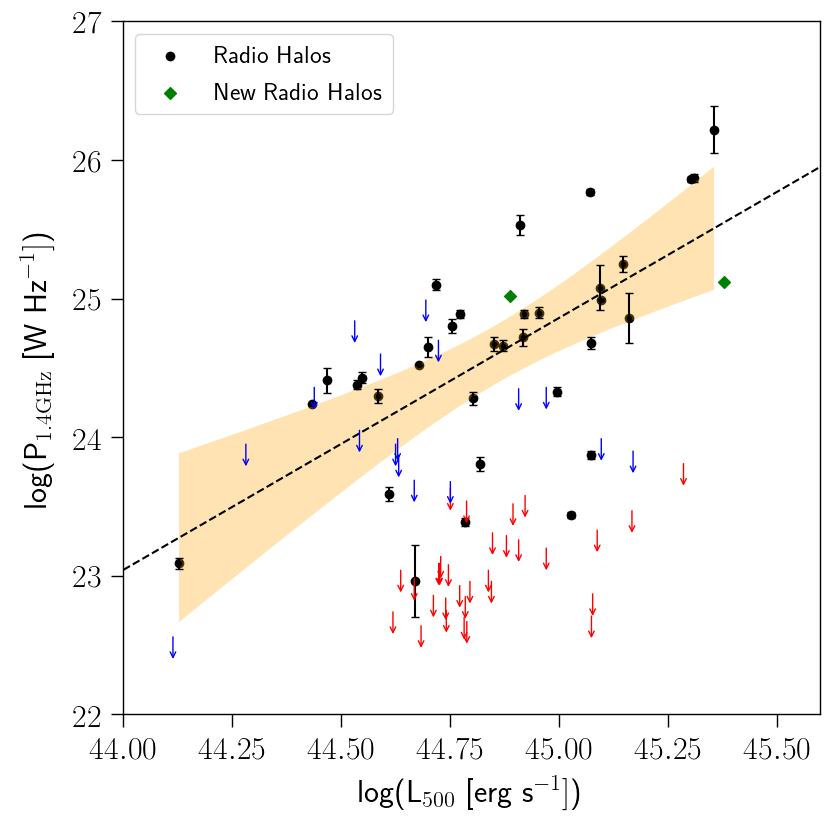}	
	\includegraphics[width=\columnwidth]{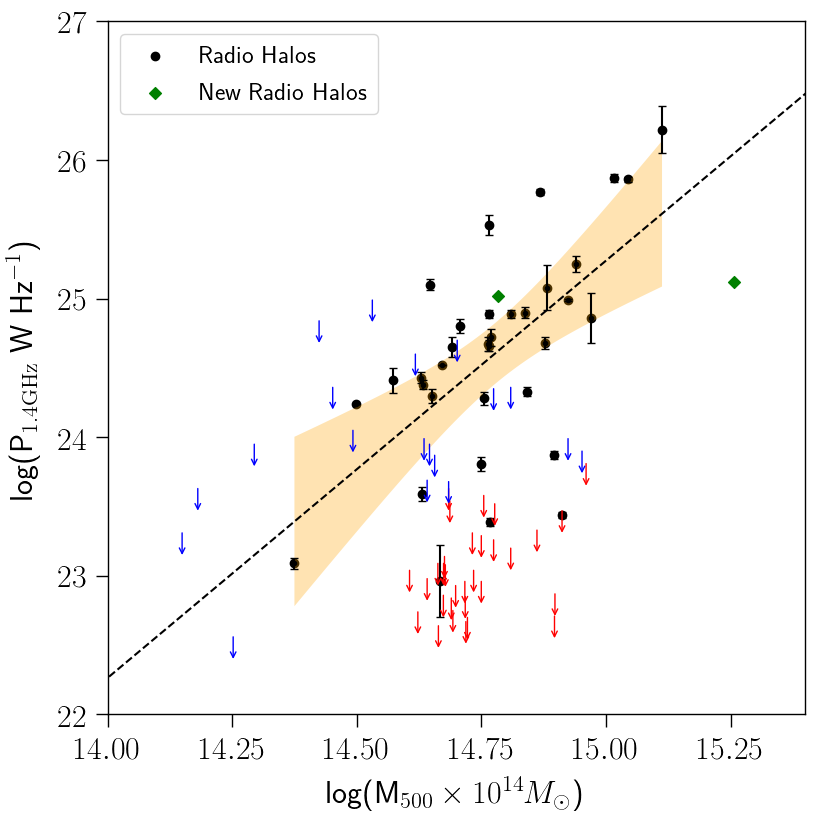}	
	\caption{The empirical relations observed in galaxy clusters hosting radio halos between the 
	X-Ray luminosity of the cluster and the radio power of the halo (\textit{Left panel}) as well as the 
	Mass of the cluster and the radio power of the halo (\textit{Right panel}).
	All previously known detections in the literature are shown in black circles while upper limits are shown 
	as red arrows.
	The two new halo detections in the paper are shown as green diamonds while the upper limits estimated
	are shown in blue arrows.
	The dashed line shows the best fit for each plot for all detections and
	the orange shaded region shows the $2\sigma$ confidence intervals.}
	\label{fig:lmp}
\end{figure*}

\noindent
In Figure~\ref{fig:lmp} we show the empirical relation between halo radio power at 1.4 GHz ($P_{1.4}$)
and the cluster X-ray luminosity ($L_{500}$) and mass ($M_{500}$). 
These studies of the relation between the radio properties of halos and the X-ray component
of galaxy clusters can provide information about the origin of the synchrotron emission from the ICM.
The radio power values for these plots were obtained from \cite{vanweeren19} and references therein
while the $L_{500}$ and $M_{500}$ values were obtained from \cite{piffaretti11}.
All the known halos from the literature are shown in black while all the upper limits from the literature
are shown in red arrows.
The best fit line was obtained following a method similar to \cite{brunetti09}.
Table~\ref{tab:lpfits} and \ref{tab:mpfits} show the best fit values for the plots using four methods:
Ordinary Least Squares (OLS) Y over X and X over Y, the bisector method and the orthogonal method.
The authors recommend using the bisector method to estimate the best fit as suggested by \cite{isobe90}
since it treats both variables symmetrically.
However, in our case \cite{piffaretti11} do not provide error values for $L_{500}$ and $M_{500}$.
Therefore, using a method that requires errors on both variables will not be appropriate.
So we use OLS (Y|X) for our best fit in the plots.

\noindent
Recently, \cite{cuciti21} performed a similar analysis on a slightly larger cluster sample where they
estimated the $M_{500}$ values themselves and thus have errors on those variables. 
They provide best fits for both the Y|X method as well as the bisector method.
Our best fit also agrees with their estimate of $2.96\pm0.5$ for the slope using the Y|X method.

\noindent
The green diamonds correspond to the two new detections from our sample.
The flux density value of the halo at 325 MHz in our images was estimated from 
the uv-subtracted images shown in Fig.~\ref{fig:j0510} and \ref{fig:j2211} 
and then scaled to 1.4 GHz (with k-correction applied) using an average halo spectral index
of $-1.3$ \citep{feretti12}.
While RXC J0510.7-0801 fits the line very well, the detected halo in the massive cluster
RXC J2211.7-0349 is almost an order of magnitude weaker in the $M_{500}-P_{1.4}$ plot.
Similar to \cite{cuciti18} who discovered two underluminous radio halos, the halo in 
this cluster could be the result of a minor merger or a result of the secondary model 
of halo formation due to hadronic collisions.
It should be noted from Fig.~\ref{fig:j2211} that the X-ray morphology of the cluster
is much more relaxed and not that disturbed as compared to RXC J0510.7-0801 in Fig.~\ref{fig:j0510}.
Thus, it is possible that the halo emission seen in the cluster is a remnant emission from 
early in the cluster's history and is now comparitively much weaker after the merger is complete.
Alternatively, the cluster could be in the earlier stages of a merger and the X-ray
emission has not been disturbed yet.

\noindent
For clusters in our sample where radio halos were not detected we tried to estimate upper limits 
(blue arrows in Fig.~\ref{fig:lmp}).
The upper limits were estimated semi-automatically.
The details of the process are described in detail in \cite{ltg20}.
Briefly, for every cluster, a model halo image with an exponential profile \citep{bonafede17} is created 
with a fixed position, redshift, size and flux density of the halo.
This image is then Fourier transformed and added to the visibility file of the cluster on a per channel basis.
The new visibility file is then imaged in \textsc{CASA} and the presence of a detection
is checked automatically.
This process is repeated for several flux densities until the halo is confirmed to be ``detected.''
The code used to estimate halo upper limits is available 
\href{https://github.com/lijotgeorge/upper-limit-calculator}{here}	
\footnote{https://github.com/lijotgeorge/upper-limit-calculator}.

\noindent
Table \ref{tab:uls} shows the estimated upper limits to halo flux densities 
at the corresponding reference frequencies as well as the extrapolated 
radio power upper limits at 1.4 GHz for other clusters in our sample. 
No limits are given for the following clusters: 
A2597, RX J0426.1+1655, RX J2237.0+1516, A2319, RXC J1308.5+5342, A0400,
A0901 and A2485.
We were unable to estimate the limits in these clusters for a few reasons as shown in Table~\ref{tab:no_uls}.
In some cases the quality of the image was extremely poor as a result of which
proper estimates could not be made while in others the extent and brightness 
of the central BCG made estimation of any upper limit in the image extremely difficult.
Note also that no upper limit has been estimated for the cluster RXJ1046.8-2535 
since no relevant X-ray information is available for the cluster.

\noindent
The GMRT Key Project on galaxy clusters was one of the first dedicated projects to make full use
of the then untested capabilities of a low frequency telescope.
The sample of clusters chosen to observe was a mix of clusters where halos had been previously detected 
at high frequencies \citep{giovannini99} and others based on discussions with X-ray astronomers.
The clusters spanned a wide range of redshifts and cluster mass so as to cover all possibilities
of radio halo detection in a then burgeoning realm of radio astronomy.

\noindent
As discussed in the previous section many of these clusters did turn out to host sources of extended emission
when they were later observed after improving the telescope's capabilities.
In this paper we showed that with modern processing techniques such as direction dependant calibration
found in SPAM it was still possible to detect the extended emission if not outright recover it.
Blind surveys like this need to be conducted even in the future
when new radio telescopes will be developed and their data needs to be carefully archived
as they can lead to new discoveries when new processing techniques are discovered.



\section{Conclusions}
\label{sec:conc}
In this work we have analysed nearly two decade old archival GMRT observations of galaxy clusters using
modern processing methods.
The GMRT Key Project on galaxy clusters was undertaken within the first few years after the 
inauguration of the GMRT. 
The clusters chosen for observation were done so after careful consideration and deliberation
with X-ray astronomers in order to make full use of the capabilities of the new telescope.

\noindent
We analysed 38 galaxy clusters and were able to detect two new radio halos in the clusters
RXC J0510.7-0801 and RXC J2211.7-0349. 
The former exhibits a highly disturbed X-ray morphology and is an example of a typical giant radio halo.
It agrees fairly well with the observed correlation between cluster luminosity and radio halo power.
RXC J2211.7-0349 on the other hand is a massive galaxy cluster that is quite relaxed 
and yet shows the presence of a weak radio halo.
This cluster does not agree with the empirical relation very well.
This could suggest that the halo emission is old and thus weaker now that there are no more sources
generating ultra-relativistic electrons in the ICM.
Or else the cluster is new and is in the very early stages of its merger
or that the halo is a result of a minor merger or due to the secondary model of halo formation.

\noindent
We also confirmed the presence of several halos and mini-halos known in the literature from the archival data.
Most of these clusters were later reobserved with the GMRT when the sensitivity of the telescope 
was much better and thus the data were of better quality. 
In the remaining 26 clusters we attempted to estimate upper limits to a possible radio halo emission.
For this purpose we developed our own semi-automated upper limit calculator.
We were able to estimate upper limits on 19 clusters in this manner. 
Upper limit estimation in the remaining clusters was not possible for various reasons.

\noindent
Blind surveys like the GMRT Key Project are crucial when a new telescope becomes operational
both to test the capabilities of the telescope as well as for the possibility of discovering new objects of 
scientific importance.
This project also shows that even legacy data with all its shortcomings can still be useful 
when new analytical techniques and algorithms are developed 
and can lead to new discoveries.

\section*{Acknowledgements}
We thank the anonymous referee for providing their valuable comments to this paper
which have helped improve it.
We acknowledge the support of the Department of Atomic Energy, Government of
India, under project no. 12-R\&D-TFR-5.02-0700. We thank the staff of the GMRT 
that made these observations possible. GMRT is run by the National Centre for Radio Astrophysics 
of the Tata Institute of Fundamental Research. 
RK acknowledges support from the DST-INSPIRE Faculty Award of the Government of India.
This research made use of Astropy,\footnote{http://www.astropy.org}
 a community-developed core Python package for Astronomy \citep{astropy:2013, astropy:2018}.
This research made use of Astroquery, an astropy affiliated package that contains 
a collection of tools to access online Astronomical data \citep{astroquery}.
This research has made use of the X-Rays Clusters Database (BAX)
which is operated by the Laboratoire d'Astrophysique de Tarbes-Toulouse (LATT),
under contract with the Centre National d'Etudes Spatiales (CNES).
This research has made use of NASA's Astrophysics Data System.

\noindent
\newline \textbf{Data Availability:} The data underlying this article were accessed from the
\href{https://naps.ncra.tifr.res.in/goa/data/search}{GMRT Online
Archive}\footnote{https://naps.ncra.tifr.res.in/goa/data/search} 
(Project IDs: 01VKK01, 02VKK01, 03VKK01, 04VKK01, 05VKK01).
The derived data generated in this research will be shared on reasonable request to the corresponding author.




\bibliographystyle{mnras}
\interlinepenalty=10000
\bibliography{vkkclusters} 


\appendix

\section{Other Clusters}

\subsection{RX J1334.3+5030}
	At a redshift of $z=0.62$ \citep{romer2000} this cluster has the highest redshift in the sample.
	It has an X-ray luminosity of $L_{\rm X} = 3.9\times10^{44}$ \eps \citep{lumb04}.
	The temperature of the cluster was estimated by \cite{kotlinin05} to be $T_{\rm X} = 4.6\pm^{0.4}_{0.3}$
	and $M_{\rm 500} = 2.73\pm^{0.45}_{0.3}\times10^{14}M_{\sun}$.

\subsection{RX J0505.3-2849}
	This cluster if at a redshift $z=0.509$ \citep{burke03} and has an X-ray luminosity of
	$L_{\rm X} = 1.1\times10^{44}$ \eps and is at a temperature of $T_{\rm X} = 2.5^{0.3}_{0.3}$ keV\citep{lumb04}.
	X-ray observations of the cluster with the $XMM-Newton$ show the presence of two peaks
	one of which has been claimed to be due to confusion with a double point source \citep{dietrich07}.

\subsection{RX J1241.5+3250}
	This cluster has a redshift of $z=0.39$ and an X-ray luminosity of $L_{\rm X} = 4.8\times10^{44}$ \eps
	\citep{romer2000}. The temperature of the cluster was estimated to be $T_{\rm X} = 6$ keV.

\subsection{RX J0318.2-0301}
	The cluster has a redshift $z=0.37 $ \citep{romer2000}. X-ray analysis of the clusters
	give the luminosity of the cluster to be $L_{\rm X} = 1.74\times10^{44}$ \eps \citep{burke03}
	and a temperature of $T_{\rm X} = 5.7^{0.3}_{0.3}$ keV \citep{ehlert09}.

\subsection{RX J1200.8-0327}
	This cluster has a redshift of $z=0.396$ \citep{mullis03} and an X-ray luminosity of
	$L_{\rm X} = 2.02\times10^{44}$ \eps \citep{vikhlinin98}. The X-ray distribution of the cluster
	is spherically symmetric and it seems to be in a relaxed state with a global temperature of
	$T_{\rm X} = 5.1\pm^{0.7}_{0.5}$ keV \citep{majerowicz04}.

\subsection{Abell 2597}
	A2597 is a nearby cool-core cluster with a redshift of $z=0.085$ \citep{strood99}.
	It has an X-ray luminosity of $L_{\rm X}=6.62\times10^{44}$ \eps \citep{reiprich02}
	and a temperature of $T_{\rm X}=4.05$ keV \citep{hudson10}.
	X-ray observations by 
	show the presence of cold, accretion flow in the cluster towards the central supermassive black hole.

\subsection{RX J0426.1+1655}
	The cluster has a redshift of $z=0.38$ and an X-ray luminosity of $L_{\rm X}=4.94\times10^{44}$ \eps
	\citep{romer2000}. The temperature of the cluster has been estimated to be $T_{\rm X}=5.4^{0.4}_{0.4}$
	keV \citep{ehlert09}.

\subsection{RX J2237.0-1516}
	This cluster is an intermediate redshift ($z=0.299$) cluster with a weak X-ray luminosity
	$L_{\rm X}=2.18\times10^{44}$ \eps \citep{romer2000}. The mean temperature of the cluster
	was estimated to be $T_{\rm X} = 3.0\pm^{0.5}_{0.5}$ keV \citep{majerowicz04}.
	\textit{XMM-Newton} analysis do not reveal any structure in the X-ray distribution \citep{majerowicz03}.

\subsection{Abell 2319}
	A2319 is a highly luminous nearby cluster with a redshift of $z=0.056$ \citep{strood99}
	and an X-ray luminosity of $L_{\rm X}=15.78\times10^{44}$ \eps \citep{reiprich02}.
	It has a temperature of $T_{\rm X} = 8.84^{0.18}_{0.14}$ keV \citep{ikebe02}.
	VLA and GBT observations of the cluster reveal the presence of a $\sim2$ Mpc size
	radio halo with a complex morphology \citep{fergiobo97,farnsworth13,storm15}.

\subsection{RX J1701.3+6414}
	This is high redshift cluster ($z=0.453$) with an X-ray luminosity of
	$L_{\rm X}=3.4\times10^{44}$ \eps \citep{lumb04} and a temperature of $T_{\rm X}=4.36^{0.46}_{0.46}$ keV
	\citep{vikhlinin09}.

\subsection{Abell 2390}
	This massive cluster has a redshift of $z=0.228$ \citep{strood99} and an X-ray luminosity of
	$L_{\rm X}=25.09\times10^{44}$ \eps \citep{allen03} and a gas temperature of
	$T_{\rm X}=8.89^{0.24}_{0.24}$ keV \citep{planck11xi}.
	The extended emission in this cool-core cluster was initially identified as a mini-halo 
	\citep{bacchi03}.
	Later observations revealed however, that the emission was much larger than previously measured
	and the source is in fact a radio halo with a steep spectrum \citep{sommer17}.

\subsection{RXC J1308.5+5342}
	This cluster has a redshift of $z=0.33$ and an X-ray luminosity of
	$L_{\rm X}=1.53\times10^{44}$ \eps \citep{lumb04}. The temperature of the cluster is
	$T_{\rm X} = 4.36^{0.38}_{0.38}$ keV \citep{ettori09}.

\subsection{RX J0847.1+3449}
	This is a high redshift cluster ($z=0.56$ \cite{vikhlinin98}) and is not very Luminous
	($L_{\rm X}=1.59\times10^{44}$ \eps \cite{lumb04}). The X-ray temperature of the cluster gas
	has also been estimated by \cite{lumb04} to be $T_{\rm X}=3.62^{0.58}_{0.51}$ keV.

\subsection{Abell 0400}
	A400 is a low redshift cluster ($z=0.024$ \cite{strood99}) with a very low X-ray luminosity
	of $L_{\rm X}=0.71\times10^{44}$ \eps and a temperature of $T_{\rm X}=2.43^{0.8}_{0.7}$ keV.
	GBT observations by \cite{farnsworth13} were not able to detect any source of extended emission 
	as the central cluster is dominated by emission due to the tailed radio galaxy 3C275.

\subsection{RX J1120.1+4318}
	This is another high redshift cluster ($z=0.6$ \cite{romer2000}). \cite{lumb04} estimated
	the X-ray luminosity of the cluster as	$L_{\rm X}=6.07\times10^{44}$ \eps and temperature
	to be $T_{\rm X}=5.45^{0.3}_{0.3}$ keV.

\subsection{Abell 1084}
	A1084 has a redshift of $z=0.132$ \citep{pimbblet06}. The cluster has an X-ray luminosity of
	$L_{\rm X}=6.82\times10^{44}$ \eps \citep{bohringer04} and a temperature of
	$T_{\rm X}=3.56^{0.5}_{0.5}$ keV \citep{pratt07}.
	Visual inspection of the ROSAT image for this cluster showed that it has an irregular morphology
	\citep{pimbblet06}.

\subsection{Abell 1689}
	A1689 is a highly luminous galaxy cluster ($L_{\rm X}=19.88\times10^{44}$) \eps \citep{reiprich02}
	at a redshift of $z=0.183$ \citep{strood99}. It has an X-ray temperature of
	$T_{\rm X}=8.17^{0.12}_{0.12}$ keV \citep{planck11xi}.

\subsection{RXC J2014.8-2430}
	This cluster has a redshift of $z=0.161$ and an X-ray luminosity of $L_{\rm X}=15.13\times10^{44}$ \eps
	\citep{bohringer04}. The temperature of the cluster is $T_{\rm X}=4.78^{0.5}_{0.5}$ keV
	\citep{pratt09}.
	A morphological analysis of the cluster \citep{weissman13} revealed the cluster
	to be in a relaxed state.

\subsection{Abell 0901}
	A901 is at a redshift of $z=0.17$ \citep{schindler00} and an X-ray luminosity of
	$L_{\rm X}=10.73\times10^{44}$ \eps \citep{bohringer04}. It also has a temperature of
	$T_{\rm X}=3.2^{0.2}_{0.2}$ keV \citep{zhang08}.
	The X-ray analysis by \cite{schindler00} revealed the cluster to have a relaxed morphology.

\subsection{Abell 0291}
	A0291 is at a redshift of $z=0.197$ \citep{strood99} and has an X-ray luminosity of
	$L_{\rm X}=6.81\times10^{44}$ \eps \citep{bohringer04}.

\subsection{RXC J1212.3-1816}
	This cluster has a redshift of $z=0.269$ and an X-ray luminosity of $L_{\rm X}=9.17\times10^{44}$ \eps
	\citep{bohringer04}.
	The cluster was observed by \cite{kale13, kale15} but no sources of extended emission were found.

\subsection{RX J1046.8-2535}
	The cluster is at a redshift of $z=0.2426$ \citep{pierre94}.
	Not much is known about this cluster either in X-ray or radio.

\subsection{Abell 2104}
	This cluster has a redshift of $z=0.153$ \citep{pimbblet06}. The X-ray luminosity of the cluster is
	$L_{\rm X}=7.26\times10^{44}$ \eps \citep{bohringer04} and a temperature of
	$T_{\rm X}=6.76^{0.19}_{0.19}$ keV \citep{baldi07}.

\subsection{Abell 2485}
	This cluster is at a redshift $z=0.247$ and has an X-ray luminosity of $L_{\rm X}=7.73\times10^{44}$
	\eps \citep{bohringer04}.
	\cite{kale13} did not find any source of diffuse emission in the cluster.

\subsection{RXC J0437.1+0043}
	This cool core cluster is at a redshift of $z=0.285$ and has an X-ray luminosity of $L_{\rm X}=11.91\times10^{44}$
	\eps \citep{ebeling00}. \cite{zhang06} estimate the temperature of the cluster to be
	$T_{\rm X}=5.1^{0.3}_{0.3}$ keV and the mass of the cluster to be $M_{\rm X} = 6.1\pm2.2\times10^{14} M_{\sun}$.
	VLA observations of the cluster \citep{feretti05} did not reveal the presence 
	of any diffuse emission in the cluster.

\subsection{Abell 2537}
	A2537 is a hot ($T_{\rm X}=8.4^{0.76}_{0.68}$ keV \cite{cavagnolo08}), highly luminous
	($L_{\rm X}=14.78\times10^{44}$ \eps \cite{cruddace02}) with a redshift of $z=0.295$ \citep{dahle02}.
	The cluster was observed as part of the GRHS \citep{venturi07}
	and the authors did not detect any source of extended emission in the cluster.
	The authors in their observations mention the emission in the cluster to be dominated
	by a tailed radio galaxy near the centre which has been detected by our analysis as well.

\begin{figure*}
	\centering
	\begin{subfigure}{\textwidth}
		\includegraphics[width=0.48\textwidth]{MCXCJ1334_3+5031_618}
		\includegraphics[width=0.48\textwidth]{MCXCJ0505_3-2849_333}
		\includegraphics[width=0.48\textwidth]{MCXCJ0505_3-2849_618}
		\includegraphics[width=0.48\textwidth]{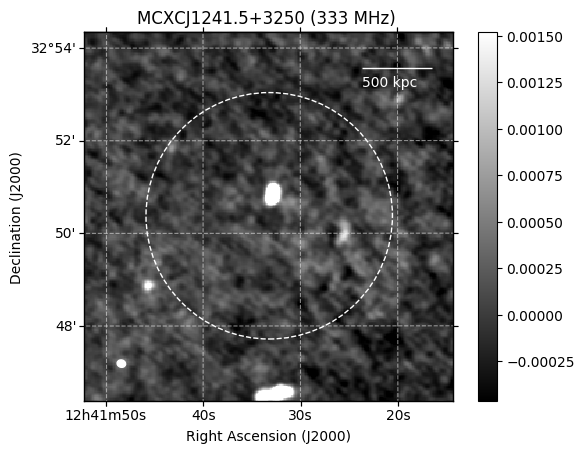}
		\includegraphics[width=0.48\textwidth]{MCXCJ1241_5+3250_618}
		\includegraphics[width=0.48\textwidth]{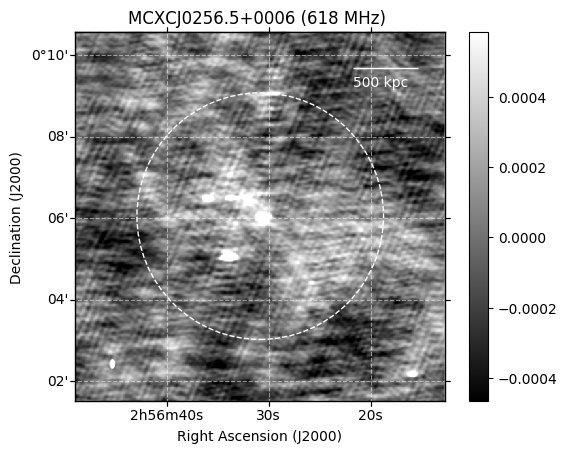}
		\subcaption{Radio images}
	 \end{subfigure}
	 \label{fig:images_1}
	 \caption{Radio images for the clusters in our sample. 
	 Dashed circles represents $R_{500}$ as given by MCXC.
	 The units of the colourbar are in Jansky/beam. 
	 These images were created using astropy with interval levels 
	 determined automatically for each image using the PercentileInterval option.
	 By default, 99\% of the pixels were used to estimate the limits with the exceptions 
	 being RXC J0847.1+3449 and RX J1046.8-2535 where 98\% of the pixels were used.}
\end{figure*}

\begin{figure*}\ContinuedFloat
	\begin{subfigure}{\textwidth}
		\includegraphics[width=0.48\textwidth]{MCXCJ0318_2-0301_333}
		\includegraphics[width=0.48\textwidth]{MCXCJ0318_2-0301_618}
		\includegraphics[width=0.48\textwidth]{A2597_618}
		\includegraphics[width=0.48\textwidth]{MCXCJ0426_1+1655_240}
		\includegraphics[width=0.48\textwidth]{MCXCJ0426_1+1655_333}
		\includegraphics[width=0.48\textwidth]{MCXCJ0426_1+1655_618}
		\caption{Radio images contd.}
	\end{subfigure}
	\label{fig:images_2}	
\end{figure*}

\begin{figure*}\ContinuedFloat
	\begin{subfigure}{\textwidth}
		\includegraphics[width=0.48\textwidth]{MCXCJ2237_0-1516_333}
		\includegraphics[width=0.48\textwidth]{MCXCJ2237_0-1516_618}
		\includegraphics[width=0.48\textwidth]{A2319_618}
		\includegraphics[width=0.48\textwidth]{A2163_317}
		\includegraphics[width=0.48\textwidth]{A2163_618}
		\includegraphics[width=0.48\textwidth]{A2345_618}
		\caption{Radio images contd.}
\end{subfigure}
\label{fig:images_3}	
\end{figure*}

\begin{figure*}\ContinuedFloat
	\begin{subfigure}{\textwidth}
		\includegraphics[width=0.48\textwidth]{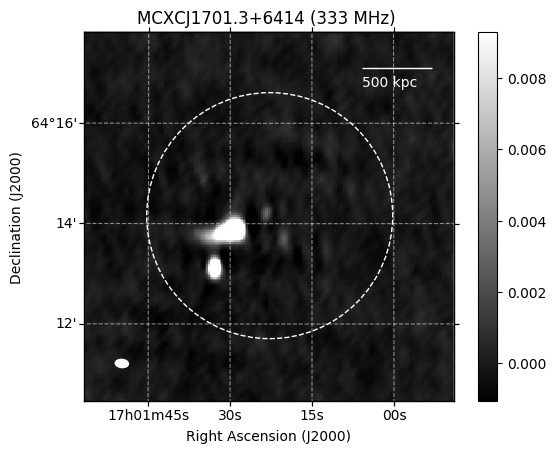}
		\includegraphics[width=0.48\textwidth]{A2390_240}
		\includegraphics[width=0.48\textwidth]{A665_240}
		\includegraphics[width=0.48\textwidth]{MCXCJ1308_5+5342_333}
		\includegraphics[width=0.48\textwidth]{MCXCJ1308_5+5342_618}
		\includegraphics[width=0.48\textwidth]{MCXCJ0847_1+3449_333}
		\caption{Radio images contd.}
\end{subfigure}
\label{fig:images_4}	
\end{figure*}

\begin{figure*}\ContinuedFloat
	\begin{subfigure}{\textwidth}
		\includegraphics[width=0.48\textwidth]{MCXCJ0847_1+3449_618}
		\includegraphics[width=0.48\textwidth]{A0400_333}
		\includegraphics[width=0.48\textwidth]{MCXCJ1120_1+4318_334}
		\includegraphics[width=0.48\textwidth]{A1084_317}
		\includegraphics[width=0.48\textwidth]{MCXCJ2211_7-0349_317}
		\includegraphics[width=0.48\textwidth]{A0545_317}
		\caption{Radio images contd.}
\end{subfigure}
\label{fig:images_5}	
\end{figure*}

\begin{figure*}\ContinuedFloat
	\begin{subfigure}{\textwidth}
		\includegraphics[width=0.48\textwidth]{A1689_317}
		\includegraphics[width=0.48\textwidth]{A0901_317}
		\includegraphics[width=0.48\textwidth]{A0291_317}
		\includegraphics[width=0.48\textwidth]{MCXCJ1212_3-1816_317}
		\includegraphics[width=0.48\textwidth]{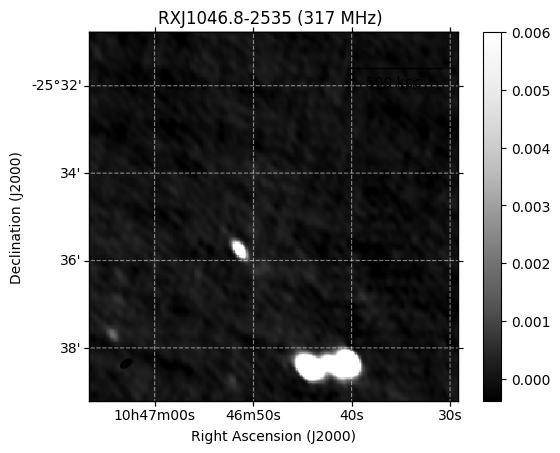}
		\includegraphics[width=0.48\textwidth]{A2104_317}
		\caption{Radio images contd. Note that no $R_{500}$ circle is shown for RXJ1046.8-2535
		since no X-ray information is available for that cluster.}
\end{subfigure}
\label{fig:images_6}	
\end{figure*}

\begin{figure*}\ContinuedFloat
	\begin{subfigure}{\textwidth}
		\includegraphics[width=0.48\textwidth]{MCXCJ0510_7-0801_317}
		\includegraphics[width=0.48\textwidth]{AS0780_317}
		\includegraphics[width=0.48\textwidth]{A2485_317}
		\includegraphics[width=0.48\textwidth]{MCXCJ0437_1+0043_317}
		\includegraphics[width=0.48\textwidth]{A0907_317}
		\includegraphics[width=0.48\textwidth]{A3444_317}
		\caption{Radio images contd.}
\end{subfigure}
\label{fig:images_7}	
\end{figure*}

\begin{figure*}\ContinuedFloat
	\begin{subfigure}{\textwidth}
		\includegraphics[width=0.48\textwidth]{A1300_317}
		\includegraphics[width=0.48\textwidth]{MCXCJ1504_1-0248_317}
		\includegraphics[width=0.48\textwidth]{MCXCJ1514_9-1523_317}
		\includegraphics[width=0.48\textwidth]{A2537_317}
		\caption{Radio images contd.}
\end{subfigure}
\label{fig:images_8}	
\end{figure*}

\bsp	
\label{lastpage}
\end{document}